\documentclass[epj]{svjour}

\begin{document}

\title{On the flavour dependence of the  
$\mathcal{O}(\alpha_s^4)$ correction to the  relation between 
running and  pole heavy quark masses}

\author{A.~L.~Kataev\inst{1,2}
\thanks{\emph{e-mail:} kataev@ms2.inr.ac.ru}
 \and 
V.~S. Molokoedov\inst{1,2,3}
\thanks{\emph{e-mail:} viktor\_molokoedov@mail.ru}
}

\institute{Institute for Nuclear Research of the Academy
of Sciences of Russia, 117312, Moscow, Russia \and 
Moscow Institute of Physics and Technology, 141700, Dolgoprudny, 
Moscow Region, Russia \and 
Landau Institute for Theoretical Physics of the Academy of 
Sciences of Russia, 142432, Moscow Region, Russia}

\date{Received: date / Accepted: date}

\abstract{
Recently the  four-loop perturbative QCD  contributions to the 
relations between pole and  running masses of charm, bottom and top quarks  were evaluated in the $\rm{\overline{MS}}$-scheme 
with identical numerical error bars.
In this
work the  flavour dependence of the $\mathcal{O}(\alpha_s^4)$ correction  to  
these  asymptotic series  is obtained in the semi-analytical form with the 
help of the least squares method.    
The numerical  structure of the corresponding asymptotic perturbative 
relations between pole and running $c$, $b$ and  $t$-quark masses
is considered  and the theoretical 
errors of the $\mathcal{O}(\alpha_s^4)$-contributions are discussed.
The explicit dependence for these relations on the renormalization scale 
$\mu^2$ and the flavour number $n_l$ is presented.     
}


\authorrunning 
\titlerunning 

\maketitle

\section{Introduction}
It is known  that    quantum chromodynamics (QCD) is the
renormalized gauge theory of quantum fields that describes strong
interactions of elementary particles and possesses the property of
confinement. As a result, it is impossible to observe quarks
in a free state. 
There are three light $u$, $d$ and $s$ quarks and
three heavy $c$, $b$ and $t$ quarks in  nature. Several theoretical
definitions of quark masses in the sectors of
light and heavy quarks are used in practice  
(see e.g. \cite{Quarksmasses}).
Among them is the notion of the  constituent  mass, which is used 
in applications of various non-relativistic quark models.
These constituent masses are not directly related to the renormalized
quark masses, which enter the QCD Lagrangian. 
The renormalized quark masses  are usually defined in the $\overline{\rm{MS}}$-scheme.
The main modern  methods  of their  determinations, including the versions
of the   QCD
sum rules  \cite{Chetyrkin:1978ta},
which were  previously used for this purpose   e.g. in 
\cite{Vainshtein:1978nn},\cite{Kataev:1982xu},\cite{Gorishnii:1983zi} and
\cite{Gorbunov:2004wy}, are  described
in the brief review \cite{Quarksmasses}.

In this work we will concentrate on the semi-analyti-cal  evaluation of the
flavour dependence of the $\mathcal{O}(\alpha_s^4)$  perturbative QCD 
correction  to
the  relation between heavy  
quark masses defined in the on-shell renormalization scheme
and their running analogues, defined 
in the 
$\overline{\rm{MS}}$-scheme. 
Since    the masses of the  bound states of light quarks
are strongly related to  various
non-pertur-bative  effects  \cite{Novikov:1981xi}, it is impossible 
to introduce for them  a  notion of pole masses, defined
in the region of high enough transferred  momentum,  where  
non-perturbative effects   are less important \footnote{In view of this the
values of the constituent heavy quark masses do not differ significantly from
the pole masses of heavy quarks.}.

The precise information about the  pole
and running heavy quark masses is  important in various phenomenological analysis.
For example, it allows to compare  theoretical QCD  prediction for the total
cross-section of the   $e^+e^-$ annihilation into hadrons process with the
experimental data, obtained  in the energy regions 
of  $\rm{J/\psi}$ and  $\rm{\Upsilon}$-mesons production   
\cite{Chetyrkin:2010ic}.
This comparison was  performed with the help of variant of the 
QCD sum rules, based on the consideration of the moments 
of the related spectral function. This 
approach was  proposed 
in \cite{Shifman:1978bx}.

The 
high-order QCD relations between running and
pole  heavy quark masses allow  also to
decrease theoretical uncertainties of the extracted from  experimental data 
Cabibbo-Kobayashi-Maskawa 
heavy quark matrix elements, and the  $V_{cb}$  element in particular.
It  enters   theoretical predictions
for the  measured  at LHCb
$B\rightarrow X_c l \bar{\nu}$ decay width.
The precise determination  of the  $b$ quark mass allows  
to perform    careful
multi-loop analysis of semileptonic decay widths of the B-meson, 
which are proportional to the  fifth
power of the $b$ quark mass \cite{Melnikov:2000qh}. 
Another important current  problem  is the accurate  determination of 
the $t$ quark mass.
The number of cosmological and particle physics
problems, related to the necessity of
decreasing theoretical uncertainties of the evaluation
of $t$ quark mass,   was 
discussed quite  recently
\cite{Bezrukov:2014ina}.
The precise  determination of   heavy quark masses, which depends  on
the knowledge of high order
perturbative QCD corrections, is 
not only the theoretically interesting calculation  task, 
but  also 
is related to  the number of 
phenomenologically important on-going
analysis of the experimental data, including the 
ones, obtained at the    LHC experiments.  

It is worth reminding  that the 
pole and running heavy quark masses are related by the asymptotic
sign-constant 
perturbative series, the nature of which is manifested in the appearance 
of the  
infrared  renormalon ambiguities 
\cite{Beneke:1994sw}, \cite{Bigi:1994em}. This leads to
the theoretical conclusion that within pure perturbation theory (PT) pole masses may be  used when   
the asymptotic structure of these series
is not manifesting itself in the  truncated perturbative relation 
between pole and running heavy quark masses.  We  will   study this important theoretical   question for  $c$, $b$ and $t$ quarks at 
the fourth-order level of PT in QCD.

\section{The  
flavour  dependence  of the $\mathcal{O}(\alpha^4_s)$ QCD 
expression for   $\overline{m}_q/M_q$: the known results}
In order to determine the ratio between the running  and pole masses of 
heavy quarks it is necessary to know the renormalisation mass constants 
in the
$\overline{\rm{MS}}$ and on-shell ($\rm{OS}$) schemes, 
which are introduced using  the notion of the 
unrenormalized bare quark mass $m_{0,q}$ and renormalized 
finite quantities $\overline{m}_q(\mu^2)$ and $M_q$  as
\begin{equation}
\nonumber 
m_{0,q}=Z^{\rm{\overline{MS}}}_m(\alpha_s)\overline{m}_q(\mu^2)~~,~~ 
m_{0, q}=Z^{\rm{OS}}_m(M^2_q, \alpha_s) M_q~
\end{equation}
Next we  consider   the ratio of the  
$\rm{\overline{MS}}$ scheme running and  pole  heavy quark masses,
namely 
\begin{equation}
\label{z_m}
z_m(\mu^2)=\frac{\overline{m}_q(\mu^2)}{M_q}=\frac{Z^{\rm{OS}}_m(M_q^2, \alpha_s(\mu^2))}{Z^{\rm{\overline{MS}}}_m(\alpha_s(\mu^2))}
\end{equation}
Here $\mu$ is the renormalization scale in the procedure of dimensional regularization \cite{'tHooft:1972fi} with $\varepsilon=(4-D)/2$. As a result of explicit manifestation of the   
multiplier $\mu^{2\varepsilon}$,  which  provides correct dimension between 
the
 bare 
and the
renormalized  QCD  coupling constant $\alpha_s(\mu^2)$ in  
the 
$\rm{\overline{MS}}$ scheme, and of the  $M^{-2\varepsilon}_q$ factor, 
which appears in the   $\rm{OS}$ scheme, the  terms  
in  $Z^{\rm{OS}}_m$  will contain the characteristic logarithms 
$L=\ln(\mu^2/M^2_q)$. However, the renormalization scale $\mu$ is a free parameter and it 
can be  fixed  as $\mu^2=M^2_q$. Fixing this normalization condition in such way we can see that all RG-governed  $L$-dependent terms   disappear and 
the expression for (\ref{z_m}) 
can be expressed through a standard QCD PT series as 
\begin{equation}
\label{ratioM}
\frac{\overline{m}_q(M_q^2)}{M_q}=z_m(M_q^2)=1+\sum\limits_{i=1}^{\infty} z^{(i)}_ma^i_s(M_q^2)
\end{equation}
where  $a_s=\alpha_s/\pi$. 
The coefficients $z^{(i)}_m$ can be represented as polynomials in  
powers of number of lighter  quarks  $n_l$ as
\begin{equation}
\label{ij}
 z^{(i)}_m=\sum\limits_{j=0}^{i-1} z^{(i,\; j)}_m n^j_l
\end{equation}
 where $n_l$=$n_f-1$ is related to the 
flavour number  
 $n_f$ of  
 the considered heavy quark.   
The first term $z^{(1)}_m$ was  calculated in  
\cite{Tarrach:1980up}. The  expression for  $z^{(2)}_m$   
was analytically evaluated  in \cite{Gray:1990yh}
and confirmed later in the process of calculations, performed in 
\cite{Avdeev:1997sz}
and \cite{Fleischer:1998dw} respectively.  The  
coefficient 
$z^{(3)}_m$ was computed in \cite{Melnikov:2000qh}
in the analytical form   and in \cite{Chetyrkin:1999qi} 
with the help of combination 
of  various semi-analytical methods. The results of these two calculations 
are in agreement with each other. According to  (\ref{ij}) the fourth coefficient  $z^{(4)}_m$ can be  expressed 
as 
\begin{eqnarray}
\label{z^4}
z^{(4)}_m=z^{(40)}_m+z^{(41)}_m n_l+z^{(42)}_m n^2_l+z^{(43)}_m n^3_l
\end{eqnarray}
The $n_l^3$ and $n_l^2$ coefficients in  (\ref{z^4})
were   computed analytically in  \cite{Lee:2013sx} and the first two terms 
are   not yet  known in
this  
form.
We will  determine them numerically using the   mathematically rigorous  
ordinary least squares (OLS) method.    

In the case of the  $\rm{SU_c(3)}$ 
gauge group with the values of the Casimir operators 
$C_F=4/3$ , $C_A=3$ and  Dynkin index $T_F=1/2$  
the 
results of the analytical calculations 
 \cite{Tarrach:1980up},\cite{Gray:1990yh},\cite{Fleischer:1998dw},\cite{Melnikov:2000qh},\cite{Lee:2013sx} 
read:
\begin{eqnarray*}
&&z^{(10)}_m=-\frac{4}{3}~ ,~~
z^{(20)}_m=-\frac{3019}{288}+\frac{\zeta_3}{6}-\frac{\pi^2\ln 2}{9}-
\frac{\pi^2}{3}~,  ~~   
z^{(21)}_m=\frac{71}{144}+\frac{\pi^2}{18}~,\\
&&z^{(30)}_m =-\frac{9478333}{93312}-\frac{61\zeta_3}{27}   
-\frac{644201\pi^2}{38880}+\frac{587\pi^2\ln 2}{162}+\frac{22\pi^2\ln^2 2}{81}+\frac{1439\pi^2\zeta_3}{432} \\ 
&&-\frac{1975\zeta_5}{216}+\frac{695\pi^4}{7776}  
+ \frac{55\ln^4 2}{162}+\frac{220}{27}\rm{Li}_4\left(\frac{1}{2}\right)~,\\ 
&&z^{(31)}_m=\frac{246643}{23328}+\frac{241\zeta_3}{72}
+\frac{967\pi^2}{648} +\frac{11\pi^2\ln 2}{81} \\ 
&&-\frac{2\pi^2\ln^2 2}{81}  
-\frac{61\pi^4}{1944}-\frac{\ln^4 2}{81}-\frac{8}{27}\rm{Li}_4\left(\frac{1}{2}\right)~,\\   
&&z^{(32)}_m=-\frac{2353}{23328}-\frac{7\zeta_3}{54}-\frac{13\pi^2}{324}~, \\ 
&&z^{(43)}_m=\frac{42979}{1119744}
+\frac{317\zeta_3}{2592}+\frac{89\pi^2}{3888}+\frac{71\pi^4}{25920}~, 
\\   
&&z^{(42)}_m=-\frac{32420681}{4478976}-\frac{40531\zeta_3}{5184}-\frac{63059\pi^2}{31104}
-\frac{103\pi^2\ln 2}{972} \\ \nonumber 
&&+\frac{11\pi^2\ln^2 2}{243}-\frac{2\pi^2\ln^3 2}{243} -\frac{5\pi^2\zeta_3}{48}+ \frac{241\zeta_5}{216}-\frac{30853\pi^4}{466560}  
\\ 
&&-\frac{31\pi^4\ln 2}{9720}  
+\frac{11\ln^4 2}{486}-\frac{\ln^5 2}{405}+\frac{44\rm{Li}_4(1/2)+24\rm{Li}_5(1/2)}{81}
\end{eqnarray*}
Here $\zeta_n$=$\sum\limits_{k=1}^{\infty}k^{-n}$ is the Riemann zeta-function, ${\rm{Li}}_n(x)=\sum\limits_{k=1}^{\infty} x^k k^{-n}$ is   the 
polylogarithmic function.     

Using these  analytical results  
we get  the following  numerical expressions 
for  the coefficients $z^{(i,j)}_m$:  
\begin{eqnarray}
\label{num}
z^{(1)}_m&=&-\frac{4}{3} \; , \; z^{(2)}_m=-14.3323+1.04136n_l \; , \; \\ 
\label{num2}
z^{(3)}_m&=& -198.706+26.9239n_l-0.65269n^2_l \; , \; \\ \label{num3}
z^{(4)}_m&=&z^{(40)}_m+z^{(41)}_m n_l-43.4824n^2_l+0.67814n^3_l 
 \end{eqnarray}
Consider now  the results of the  
recent   
complicated  numerical computer  calculations \cite{Marquard:2015qpa}  
of the fourth  
coefficient $z^{(4)}_m$ at fixed values of  $n_l$, namely 
\begin{eqnarray}
\label{z^4_m3}
z^{(4)}_m \bigg\vert_{n_l=3}&=&-1744.8\pm 21.5 , \\ \label{z^4_m4}
z^{(4)}_m \bigg\vert_{n_l=4}&=&-1267.0\pm 21.5 , \\ \label{z^4_m5}
 z^{(4)}_m \bigg\vert_{n_l=5}&=&-859.96\pm 21.5
\end{eqnarray}
where  $\sigma_{n_l}$=21.5 are   related to   
the uncertainties of   computations 
of  massive four-loop   on-shell propagator master  integrals,
which  enter into the procedure of evaluation of (\ref{z^4}) at fixed $n_l$ 
\cite{Marquard:2015qpa}.  
 
 Note that the presented 
inaccuracies  in (\ref{z^4_m3})-(\ref{z^4_m5}) are equal to each other and 
do not depend on $n_l$. In view of this  surprising from the first 
glance feature it is worth to describe how they were fixed in the work 
of \cite{Marquard:2015qpa}. In the process of these computations it was necessary to evaluate   386 on-shell master integrals.
However, only 54 integrals 
were  calculated analytically, while the rest of them  were  computed 
numerically by means of the  FIESTA program, developed in  
\cite{Smirnov:2008py},\cite{Smirnov:2009pb},\cite{Smirnov:2013eza}.
This program expresses the   
results for the  integrals in the form of their  $\epsilon$-expansion
with the  numerically evaluated  coefficients  and definite numerical errors. These errors 
are  interpreted as   standard deviations 
and are combined quadratically  in the uncertainties for physical result. 
The final value of  $\sigma_{n_l}$  are defined in  \cite{Marquard:2015qpa}
by multiplying these uncertainties   by the factor five (!?). 

From our point of view the $n_l$-independence 
of the inaccuracies  $\sigma_{n_l}$=21.5 can 
be explained by the fact that these errors   
are  almost entirely defined by the  
error of the 
constant term $z^{(40)}_m$, which 
is determined by the  set of  four-loop diagrams without 
insertion of the fermion loops into gluon propagators, whereas the uncertainties of   
$n_l$-dependent $z^{(41)}_m$-term are negligible. A possible further study of the reliability 
of this statement   
may   clarify whether the described above feature of  numerical 
calculations performed  
by the authors of  \cite{Marquard:2015qpa}  is really  
$n_l$-independent.

\section{The determination of the analytically unknown four-loop 
contributions by  the least squares method}  
Let us use
the presented in (\ref{z^4_m3})-(\ref{z^4_m5}) numerical results 
to determine the values  of the first two analytically unknown  
coefficients $z^{(40)}_m$ and 
$z^{(41)}_m$ in (\ref{z^4})
by means of the  ordinary  least squares (OLS) method.
This strict mathematical  
method is known as a standard approach for  solution 
of the  overdetermined systems of linear equations 
and allows to determine  errors of the obtained results. 

In our case we have overdetermined system of three 
linear equations with two unknown 
coefficients $z_m^{(40)}$ and $z^{(41)}_m$. 
Combining equation (\ref{num3}) with the numerical results 
of (\ref{z^4_m3})-(\ref{z^4_m5}) we get 
\begin{eqnarray}
\nonumber 
z^{(40)}_m+3z^{(41)}_m&=&-1371.77 , \\ \label{system}
z^{(40)}_m+4z^{(41)}_m&=&-614.68 , \\ \nonumber 
z^{(40)}_m+5z^{(41)}_m&=&142.32 
\end{eqnarray}
Within the  OLS method 
one should define the following  
residuals
$\Delta_{l_k}=z^{(40)}_m+z^{(41)}_m n_{l_k}-y_{l_k}$,
where   index  $1\leq k\leq 3$ denotes the  number of the concrete   
equation in the   system
(\ref{system})  
and $y_{l_k}$  are the numbers, given in  the r.h.s. of these   equations, 
which are
determined as $y_{l_k}=z^{(4)}_{m_k}-z^{(42)}_mn^2_{l_k}-z^{(43)}_mn^3_{l_k}$, where  
$z^{(4)}_{m_k}$     
 is the  one from the calculated in \cite{Marquard:2015qpa}     three   
concrete expressions  for  $z^{(4)}_m$ at fixed number of $n_l$ 
(see (\ref{z^4_m3})-(\ref{z^4_m5})) and  
$z^{(42)}_m$ and $z^{(43)}_m$ are the  known coefficients, 
which enter (\ref{num3}). 

The second 
important ingredient of the   OLS method   is the  
characteristic function, determined by  the sum of squared  
residuals    
\begin{equation}
\label{Phi}
\Phi(z^{(40)}_m, z^{(41)}_m)=\sum\limits_{k=1}^3 \Delta_{l_k}^2
\end{equation}    
The solution   $({z^{(40)}_m , z^{(41)}_m})$  of the presented 
system    exists and is defined  uniquely.  
 Indeed,  the function  $\Phi(z^{(40)}_m, z^{(41)}_m)$ always  has the 
minimum, determined from   the following equations   
\begin{equation}
\label{OLSeq}
\frac{\partial\Phi}{\partial z^{(40)}_m}=0 , \;\; \frac{\partial \Phi}{\partial z^{(41)}_m}=0 
\end{equation}
These conditions allow us to find the  numerical values for 
the coefficients $z^{(40)}_m$ and  $z^{(41)}_m$.
Within the  OLS method it is also possible to 
define for them  
the following theoretical   uncertainties 
\begin{eqnarray}
\label{error1}
\Delta z^{(40)}_m&=&\sqrt{\sum\limits_{k=1}^3 \left( \frac{\partial z^{(40)}_m}{\partial y_{l_k}} \Delta  y_{l_k} \right)^2}=  
\frac{\sqrt{\sum\limits_{k=1}^3 n^2_{l_k}}}{\sqrt{3\sum\limits_{k=1}^3 n^2_{l_k}-\left(\sum\limits_{k=1}^3 n_{l_k}\right)^2}}\Delta y_l~,
\end{eqnarray}  
\begin{eqnarray}
\label{error2}
\Delta z^{(41)}_m&=&
\sqrt{\sum\limits_{k=1}^3 \left( \frac{\partial z^{(41)}_m}{\partial y_{l_k}} \Delta  y_{l_k} \right)^2}=
\frac{\sqrt{3}\Delta y_l}{\sqrt{3\sum\limits_{k=1}^3 n^2_{l_k}-\left(\sum\limits_{k=1}^3 n_{l_k}\right)^2}}  
\end{eqnarray}
where for each  k=1, 2, 3 the values
$\Delta y_{l_k}$ $\equiv$ $\Delta y_l= \sigma_{n_l}=21.5$. It is worth emphasizing that errors of the considered by us OLS method can not be 
eliminated and in addition to the uncertainties, given  in (\ref{z^4_m3})-(\ref{z^4_m5}),
it is necessary to take them into account. 

The determined by (\ref{OLSeq}) numerical  values of $z^{(40)}_m$ and $z^{(41)}_m$ coefficients   
with the fixed by (\ref{error1}) and (\ref{error2}) corresponding theoretical uncertainties
read:
\begin{equation}
\label{answer+error}
z^{(40)}_m =-3642.9\pm 62.0 , \;\; z^{(41)}_m=757.05\pm 15.20 
\end{equation}
In contrast to the results  (\ref{z^4_m3})-(\ref{z^4_m5}) of  
\cite{Marquard:2015qpa}, where in accordance with our guess, 
presented above,  all   uncertainties $\sigma_{n_l}$ in the determination of 
$z^{(4)}_m$ at fixed $n_l$ may be associated with the errors of $z^{(40)}_m$-term, 
in our results inaccuracies are found not only in the  $z^{(40)}_m$-term, 
but in $n_l$-dependent contribution as well (though it is 4 times smaller). 
  
In the case of applications of the OLS method  theoretical  error of  
$z^{(40)}_m$ is 
almost three times larger than the  one, 
presented in \cite{Marquard:2015qpa}, 
and the OLS   uncertainty of $z^{(41)}_m$-term  is comparable with $\sigma_{n_l}$.

Theoretical uncertainties in (\ref{answer+error})  were computed using only  
{\it three}  available  from  (\ref{system}) points, 
which  form  a triangle on the plane in coordinates
$(n_l;y_l)$. In view of this  the  errors in (\ref{answer+error}) 
may be  
overestimated \footnote{More detailed clarification of this statement is 
given in Note added at the end of the paper.}.
In our studies we do not  
consider a correlation of these   
{\it  three}  data points. Indeed, the initial   
quadratic uncertainties $\sigma_{n_l}$
does not exceed 10-15 $\%$ of the r.h.s. expressions  in   (\ref{system}). 
The resulting numbers are 
small.   Therefore we  neglect  the consideration of the possible correlation of the errors in our final result (\ref{answer+error}).
 
Note, that there  is  a  
criterion of the quality of the application of the  OLS  method. It presumes the   evaluation 
of the coefficient  $r$, which is  defined as the geometric mean of regression coefficients. 
In our case it has the following form 
\begin{eqnarray*}
r&&=\sqrt{\rho_{n_ly_l}\rho_{y_ln_l}}=  
\frac{3\sum\limits_{k=1}^3 n_{l_k}y_{l_k}-\sum\limits_{k=1}^3 n_{l_k}\sum\limits_{k=1}^3 y_{l_k}}{\sqrt{\left(3\sum\limits_{k=1}^3 n^2_{l_k}-\left(\sum\limits_{k=1}^3 n_{l_k}\right)^2\right)\left(3\sum\limits_{k=1}^3 y^2_{l_k}-\left(\sum\limits_{k=1}^3 y_{l_k}\right)^2\right)}}
\end{eqnarray*}
In the case when $r$=1 the function  $y_l(n_l)$ has  a precise linear  dependence
on $n_l$.  In our case we have $r$=0.9999. 
This means that even in the case 
of three equations the OLS method  is valid,  
and gives  rather realistic 
results 
with the related to    
theoretical 
errors,  which, however, are becoming more  realistic in case when it 
is used the OLS method for solving more than {\it three} initial points (see 
Note added).  
 
\section{Results and discussions}

Taking into account the presented in (\ref{num})-(\ref{num3}) 
numerical 
expressions of  the available results of analytical calculations 
and the  OLS based expression from (\ref{answer+error}), we arrive to the 
following $\mathcal{O}(\alpha_s^4)$ relation between pole and 
the
$\rm{\overline{MS}}$-scheme  
running 
heavy  quark masses:     
\begin{eqnarray}
\nonumber
&&M_q \approx \overline{m}_q(M^2_q)
(1+1.3333a_s( M^2_q) \\ \nonumber 
&&+(-1.0414n_l+16.110)a^2_s( M^2_q) \\ \label{result3}
&&+(0.6527n^2_l-29.701n_l+239.30)a^3_s( M^2_q) \\ \nonumber  
&&+(-0.6781n^3_l+46.309n^2_l \\ \nonumber 
&&-(864.25\pm 15.20)n_l+(4457.6\pm 62.0 )a^4_s( M^2_q))  \nonumber 
\end{eqnarray}
\begin{table*}[t]
  \centering
\begin{tabular}{|c|c|}
\hline
$n_l $ & $\textbf{ The QCD  $\mathcal{O}(\alpha_s^4)$ relations of 
$M_q$ to  the running masses  $\overline{m}_q(M^2_q)$  }$    \\
\hline
$3$ & $M_c\approx\overline{m}_c(M^2_c)( 1+1.3333a_s(M^2_c)+12.985a^2_s(M^2_c)+156.07a^3_s(M^2_c)+(2263.4\pm 76.9)a^4_s(M^2_c)) $  \\
\hline
$4$ & $M_b\approx\overline{m}_b(M^2_b)( 1+1.3333a_s(M^2_b)+11.944a^2_s(M^2_b)+130.93a^3_s(M^2_b)+(1698.2\pm 86.8)a^4_s(M^2_b)) $  \\
\hline
$5$ & $M_t\approx\overline{m}_t(M^2_t)( 1+1.3333a_s(M^2_t)+10.903a^2_s(M^2_t)+107.11a^3_s(M^2_t)+(1209.4\pm 98.0)a^4_s(M^2_t)) $  \\
\hline
$ $ & $\textbf{  The QCD  $\mathcal{O}(\alpha_s^4)$ relations of 
$M_q$ to  the running masses  $\overline{m}_q(\overline{m}^2_q)$  }$    \\
\hline
$3$ & $M_c\approx\overline{m}_c(\overline{m}^2_c)( 1+1.3333a_s(\overline{m}^2_c)+10.318a^2_s(\overline{m}^2_c)+116.49a^3_s(\overline{m}^2_c)+(1691.1\pm 76.9)a^4_s(\overline{m}^2_c)) $  \\
\hline
$4$ & $M_b\approx\overline{m}_b(\overline{m}^2_b)( 1+1.3333a_s(\overline{m}^2_b)+9.277a^2_s(\overline{m}^2_b)+94.41a^3_s(\overline{m}^2_b)+(1224.0\pm 86.8)a^4_s(\overline{m}^2_b)) $  \\
\hline
$5$ & $M_t\approx\overline{m}_t(\overline{m}^2_t)( 1+1.3333a_s(\overline{m}^2_t)+8.236a^2_s(\overline{m}^2_t)+73.63a^3_s(\overline{m}^2_t)+(827.3\pm 98.0)a^4_s(\overline{m}^2_t)) $  \\
\hline
\end{tabular}
\vspace{0.2cm}\\
{Table 1. The PT QCD relations between  pole and  
the  $\rm{\overline{MS}}$-scheme 
running $c$, $b$ and $t$ quark masses  
 for two normalization scales.}
\end{table*} 
At present it is commonly accepted 
to present  the values of the   running heavy quark masses  fixed  
at the renormalization scale   
$\mu^2=\overline{m}^2_q$.   
 At the four-loop level they are related to   the heavy quark pole  masses 
by  the following perturbative  expression  
\begin{equation}  
\label{l}
M_q=\overline{m}_q(\overline{m}^2_q)\bigg(1+ \sum\limits_{i=1}^4 l_i 
a_s^i(\overline{m}^2_q)\bigg)~~~.
\end{equation}
The coefficients $l_i$ 
are determined  
using the RG-based equations, which were used in \cite{Kataev:2010zh} 
at the three-loop level    
and  are  presented  in the Appendix at the four-loop level.  
Their  numerical expressions 
have the following form:
\begin{eqnarray}
\label{l1-l2}
l_1&=&\frac{4}{3}~, ~~ l_2=13.4433-1.04136n_l~, \\ \label{l3}
l_3&=&190.595-26.6551n_l+0.65269n^2_l~, \\ \label{l4}
l_4&=&-86.54-z^{(40)}_m+(11.221-z^{(41)}_m)n_l \\ \nonumber 
&+&43.3962n^2_l-0.67814n^3_l 
\end{eqnarray}
where the obtained in this work OLS  values for   
$z^{(40)}_m$ and $z^{(41)}_m$  are presented in  (\ref{answer+error}).
Taking them into account we get the following expression  for the  $l_4$-term:
\begin{eqnarray}
\label{l_4}
l_4&=&(3556.4 \pm 62.0) - (745.83 \pm 15.2)n_l \\ \nonumber 
&+&43.396n^2_l -0.6781n^3_l
\end{eqnarray}
Here we emphasize again that the resulting uncertainties are  
contained not only in the  $n_l$ independent contribution, but in proportional 
to $n_l$ coefficient as well. 
Our result agrees rather well   with the    independently   obtained 
in  \cite{Kiyo:2015ooa}   
by
another method  
following expression     
\begin{eqnarray}
\label{l_4K}
l_4&=&(3556.5 \pm 21.5)- 745.85n_l \\ \nonumber 
&+&43.396n^2_l -0.6781n^3_l~~~
\end{eqnarray}
 Both methods of the  determination 
of  the $n_l$-dependence of $l_4$  
have  common 
features. They are using 
the same input, namely, the   
results of the performed in \cite{Lee:2013sx}
analytical calculations 
of the $n_l^2$ and $n_l^3$  $\mathcal{O}(\alpha_s^4)$
coefficients $z_m^{(42)}$ and $z_{m}^{(43)}$ 
in  the ratio between  the $\rm{\overline{MS}}$-scheme running and 
pole heavy quark masses.   
Next, 
both methods are using 
the
obtained in \cite{Marquard:2015qpa} 
numerical expressions 
(\ref{z^4_m3})-(\ref{z^4_m5}) 
of the  whole values of the four-loop corrections to 
considered ratios. 

However, these two methods are completely different 
from theoretical   point of view. Firstly,    
in our work we use 
rather rigorous OLS mathematical method, while  the approach of  
\cite{Kiyo:2015ooa}  is  based on 
application of less 
mathematically motivated  
special fitting procedure, 
which is supplemented by 
the extra 
{\it theoretical}   information, namely 
by the derived   
in \cite{Beneke:1994qe} and \cite{Beneke:1994sw}    
renormalon-based large $\beta_0$-representation of the 
$n_l$-dependence for the  $l_4$ coefficient  
\footnote{It is 
known from \cite{Broadhurst:2002bi} that    
the application of this approach for other physical 
quantities gives reasonable prediction of the $n_l$-dependence 
for the  three-loop perturbative QCD approximations.}.  
In our studies it was not necessary to  use   
this additional  theoretical input.  

Secondly, we apply  the OLS method to obtain  
the numerical values of $z_{m}^{(40)}$ and $z_m^{(41)}$ contributions 
to (\ref{z^4}), and  after this we get the corresponding  numerical expression for $l_4$ (see (\ref{l4}) and (\ref{l_4})) using 
the appropriate RG-equations.  
In the work of \cite{Kiyo:2015ooa} the analogous expression for $l_4$, 
which is given in (\ref{l_4K}), was obtained as the result of application 
of the fitting procedure to the analytical contributions  \cite{Lee:2013sx}  and numerical results of     \cite{Marquard:2015qpa}, which  
are related to the $n_l$-dependence of $l_4$-coefficient  directly.  
In view of these two arguments   the coincidence of the central values of the $n_l$-independent and 
the proportional to the first power of $n_l$ terms in our result  (\ref{l_4}) and in  the 
similar result  (\ref{l_4K}) from     \cite{Kiyo:2015ooa}
is the non-trivial fact and gives extra 
confidence in the validity of both  
mathematical OLS method and this     
more physical fitting procedure. 
 In addition the OLS based result (\ref{l_4}) 
has   
the sign-alternating structure of the contributions which are 
proportional to the  powers of  $n_l$.   
A posteriori this feature of the OLS based results   
supports the applied in  \cite{Kiyo:2015ooa}   
renormalon-based  large-$\beta_0$ theoretical  considerations. 

The numerical $\mathcal{O}(\alpha_s^4)$  approximations 
of the relations between pole and  $c$-,  $b$- and $t$-quarks  
running masses
are presented in Table 1. The results of Table 1 demonstrate  that the general  asymptotic structure 
of the perturbative QCD series really manifest itself. Indeed, one can see that
all relations contain sign-constant and significantly  growing coefficients
of the corresponding PT series. Moreover, the table demonstrates 
the importance of the  four-loop QCD  contributions  
in all given above relations.

Note, that    
the  
OLS mathematical 
method allows 
to formalize the procedure of fixing    
theoretical error bars of the
$\mathcal{O}(\alpha_s^4)$ coefficients.
In the presented in Table 1
relations
 the obtained by OLS method  uncertainties 
 are  based on given above equations (\ref{error1}) and (\ref{error2}). 
As it was already mentioned previously these errors are 
using only  {\it three} available from the system  (12)  initial points   
and may be overestimated.  

In the result  of applications of the fitting 
procedure of  
\cite{Kiyo:2015ooa}  the error of $n_l$-independent  term in (\ref{l_4K}) 
is fixed by  the error of the numerical calculations of 
\cite{Marquard:2015qpa}.  

Let us now study the concrete behaviour of 
the
QCD PT   $\mathcal{O}(\alpha_s^4)$ relation between
pole and the  $\rm{\overline{MS}}$-scheme running  
masses of heavy quarks. 
In our numerical studies 
we will  use  the  world average values of the  running masses of 
the $c$ and $b$-quarks, which are 
given in the  review of particle  physics 
properties volume \cite{Agashe:2014kda}, namely 
$\overline{m}_c(\overline{m}^2_c)$=1.275 $\rm{GeV}$ and $\overline{m}_b(\overline{m}^2_b)$=4.180 $\rm{GeV}$. 
The value of the running   $t$ quark mass 
$\overline{m}_t(\overline{m}^2_t)$=163.643 $\rm{GeV}$
is taken from the work \cite{Marquard:2015qpa}.  It 
does not contradict the results presented in PDG, which can be extracted 
from measurement of $\sigma(t\bar{t})$ in $p\bar{p}$ collisions.

The  
expression for $\alpha_s$ is 
 defined through the expansion in inverse powers of 
 ${\rm{L}}=\ln(\overline{m}^2_q/
{\rm{\Lambda}}^{(n_f)2}_{\overline{\rm{MS}}})$ terms with the parameters 
${\rm{\Lambda}}^{(n_f)}_{\overline{\rm{MS}}}$, which depend on the flavour number of  quarks  
($n_f=n_l+1$) 
and the  order of approximation of the QCD $\beta$-function in the 
$\overline{\rm{MS}}$-scheme. 
For the  $b$ quark we take the average world value of 
${\rm{\Lambda}}^{(n_f=5)}_{\overline{\rm{MS}}}$ from 
\cite{QCDPDDG}, which is consistent with the world average 
 value $\alpha_s(M^2_Z)=0.1185$. 
In order to obtain values ${\rm{\Lambda}}^{(n_f=4)}_{\overline{\rm{MS}}, \; \rm{N^3LO}}$ 
and ${\rm{\Lambda}}^{(n_f=6)}_{\overline{\rm{MS}}, \; \rm{N^3LO}}$ we use the 
$\rm{N^3LO}$  matching transformation conditions from
\cite{Chetyrkin:1997sg}. 
The 
corresponding results read:
\begin{eqnarray}
\label{Lnf4}
{\rm{\Lambda}}^{(n_f=4)}_{\overline{\rm{MS}}, \; \rm{N^3LO}}&=&297~{\rm{MeV}}, \;\;\;   \alpha_s^{\rm{N^3LO}}(\overline{m}^2_c) \approx 0.399 \\ \label{Lnf5}
{\rm{\Lambda}}^{(n_f=5)}_{\overline{\rm{MS}}, \; \rm{N^3LO}}&=&215~{\rm{MeV}}, \;\;\;   
\alpha_s^{\rm{N^3LO}}(\overline{m}^2_b)\approx 0.227 , \\ \label{Lnf6}
{\rm{\Lambda}}^{(n_f=6)}_{\overline{\rm{MS}}, \; \rm{N^3LO}}&=&91~{\rm{MeV}}, \;\;\;\;\; \alpha_s^{\rm{N^3LO}}(\overline{m}^2_t) \approx 0.109
\end{eqnarray}
The obtained   results  for 
${\rm{\Lambda}}^{(n_f=4)}_{\overline{\rm{MS}}}$ and 
${\rm{\Lambda}}^{(n_f=6)}_{\overline{\rm{MS}}}$ are in agreement 
with the ones, given in \cite{QCDPDDG}. This gives us confidence 
that the presented above N$^3$LO expressions for  
$\alpha_s$ at different scales are consistent with the  world average value $\alpha_s(M^2_Z)$.

Using the given in Table 1   QCD  $\mathcal{O}(\alpha_s^4)$ relations of 
$M_q$ to  the fixed above values of the  running masses   $\overline{m}_q(\overline{m}^2_q)$ 
and the results  for  $\alpha_s$ from (\ref{Lnf4})-(\ref{Lnf6}) 
we get the following numerical expressions for the four-loop perturbative 
series we are interested in  
\begin{eqnarray}
\nonumber 
&&\frac{M_c}{1~\rm{GeV}}\approx  1.275+0.216+0.213 \\ \label{Mc}
&&+0.305+0.563\pm 0.026~,  \\ \nonumber 
&&\frac{M_b}{1~\rm{GeV}}\approx  4.180+0.403+0.202 \\ \label{Mb} 
&&+0.149+0.140\pm 0.010~, \\ \nonumber 
&&\frac{M_t}{1~\rm{GeV}}\approx  163.643+7.549+1.613 \\ \label{Mt}
&&+0.499+0.194\pm 0.023=173.498 \pm 0.023
\end{eqnarray}
where the theoretical OLS inaccuracies of the $b$ and  $t$ quark pole masses 
are 2.5 and 4.6 times larger  than  the  
errors, presented in \cite{Marquard:2015qpa}.
Indeed, the OLS errors include the uncertainties of \cite{Marquard:2015qpa}, 
given in (\ref{z^4_m3})-(\ref{z^4_m5}), as 
a part of the determination of the   theoretical inaccuracies  with the help  
of the OLS method.

All numerical  corrections give  a significant contributions to the 
expressions for the pole heavy quark masses. Moreover, 
in the case of $c$ quark, the asymptotic nature of PT series  is  manifesting  
itself from the  third order of PT. Indeed, 
the numerical values of the  fourth and fifth terms are larger than the third 
term, which corresponds to the next-to-leading $\mathcal{O}(\alpha^2_s)$ term. 
In view of this it is really impossible to fix  the value of the pole $c$ 
quark mass at the fourth 
and even third level of perturbative QCD.   
In the case of the  
$b$ quark  the numerical value  of the fourth order term is comparable 
with the  $\mathcal{O}(\alpha_s^3)$ contribution. 
These features demonstrate that the 
studied  theoretically in \cite{Beneke:1994sw},
\cite{Bigi:1994em} IR renormalon long-distance  contributions to the 
PT series for the $c$ and $b$ quark pole  masses are manifesting themselves 
rather early, namely at the third  and  fourth order of corresponding   
perturbative series.
The expression (\ref{Mc}) 
clarifies  the known conclusion  why instead of    
the pole $c$-quark mass  it is commonly accepted to use 
the running $c$-quark mass in the number of the concrete phenomenological 
applications. For  the $b$-quark the concept of the 
pole mass may be still applicable at the  
truncated $\mathcal{O}(\alpha_s^3)$ perturbative analysis. 
However,  in view of the manifestation 
of the asymptotic structure of the PT series of (\ref{Mb}) at the four-loop 
level it is indeed more rigorous to use  the running $b$ quark mass 
in the related high-order perturbative QCD  phenomenological 
studies.

In the case of the $t$-quark  mass  
the evaluated   PT QCD  corrections are 
decreasing. 
However, the effect of $\mathcal{O}(\alpha^4_s)$ correction is not negligible. 
Its uncertainty was  fixed within  the 
OLS approach. 
The accuracy of our method turns out to be 23 $\rm{MeV}$, which is over 5 
times larger than the similar  theoretical uncertainty, estimated 
in \cite{Marquard:2015qpa}. 
The difference may be important in the 
detailed considerations of the theoretical studies, discussed in 
\cite{Bezrukov:2014ina}. The clarification of the  
raised in our work  problems, 
related to the determination of real  precision of the four-loop QCD 
contribution to the relation between  pole and running 
$t$-quark masses  are also important in view 
of the existence of the electro-weak (EW) and mixed EW-QCD corrections 
to the pole-running  $t$-quark mass relation  
\cite{Jegerlehner:2012kn}. 
These corrections are  comparable with the expressions for  
the four-loop QCD contributions.        
The discussed in this work   theoretical 
uncertainties can be removed  after direct analytical 
calculation of  the $z^{(40)}_m$ and $z^{(41)}_m$ coefficients 
in (\ref{z^4}). The  preparations for these 
calculations have  already started  \cite{Lee:2015eva} 
from the creation of the first computer program.

\section{Conclusion}
In this work we determine the constant term $z^{(40)}_m$
and the coefficient  $z^{(41)}_m$   
of the   flavour dependent $\mathcal{O}(\alpha_s^4)$  contribution to the  
ratio  
$\overline{m}_q(M_q^2)/M_q$ 
by the mathematical  least squares method (which is not  related to any fitting 
procedure)  and 
evaluate the inaccuracies  of these two coefficients using 
the obtained  in \cite{Marquard:2015qpa} {\it three} 
given in (\ref{z^4_m3})-(\ref{z^4_m5})
numerical expressions.
In this case  
the  fixed by the OLS  method whole uncertainties  of the  four-loop 
corrections to the relation between pole and running heavy quark masses 
turn out to be 6.5, 2.5 and 4.6 
times larger   than 
the   estimated 
in   similar    
errors  for the  $c$, $b$ and $t$ quark  masses respectively. 
Theoretical  arguments in favour 
of the  applicability of the OLS  method 
for the mathematically consistent determination of the 
central numerical values of  two contributions to the $n_l$-dependent 
expression for the  four-loop correction to the relations  between pole and  
the $\rm{\overline{MS}}$-scheme running  
heavy quark masses are presented. The asymptotic structure of these 
perturbative relations     
It will be interesting to understand   
the reason of differences of the obtained 
in this work 
error-bars for the $\mathcal{O}(\alpha_s^4)$  contribution to the  
ratio between  running and pole 
heavy quark masses from  the ones 
obtained as the result of the performed in  \cite{Marquard:2015qpa} 
important numerical calculations.  
The
necessity of 
the direct analytical calculation of  $z^{(40)}_m$ and  $z^{(41)}_m$  
terms  is emphasized.\\

{\bf Acknowledgments}\\

We are grateful to M.Y. Kalmykov, R.N. Lee, V.A. Smirnov
and especially I. Masina for useful discussions.  
The work of A.K. is  done within the scientific program  of the 
Russian Foundation 
for Basic Research project no.14-01-00695. The work of V.M., related to 
the studies the asymptotic structure of 
the four-loop relation between pole and running 
$c$- and $b$-quark masses,  is done within the scientific program of the 
Russian Science Foundation grant  No.16-12-10151.\\

{\bf Appendix}\\

Here we describe in details how to obtain the flavour-dependent  
 $\mathcal{O}(\alpha_s^4)$   
relation between pole and $\rm{\overline{MS}}$-scheme  running  masses of heavy quarks, normalized at the non-fixed normalization scale $\mu^2$. 
To solve this 
problem we use the four-loop approximation of the QCD $\beta$-function 
in the $\rm{\overline{MS}}$-like schemes, which is defined as  
\begin{eqnarray}
\label{beta}
\beta(a_s)&=&\mu^2\frac{\partial a_s(\mu^2)}{\partial\mu^2}=-\sum\limits_{i=0}^{3} \beta_i a_s^{i+2}~~~~
\end{eqnarray}   
The numerical expressions of the coefficients $\beta_i$, 
which are expanded in powers of  
$n_l=n_f-1$,   have the 
following form 
\begin{eqnarray*}
&&\beta_0= 2.5833-0.16666n_l~, \\
&&\beta_1=5.5833-0.79166n_l~, \\
&&\beta_2=18.045-4.1808n_l+0.09403n^2_l~,\\
&&\beta_3=88.684-23.951n_l+1.5999n^2_l+0.00585n^3_l
\end{eqnarray*}
The  scheme-independent coefficients $\beta_0$ and $\beta_1$ were 
analytically evaluated in \cite{Gross:1973id},\cite{Politzer:1973fx}
and \cite{Jones:1974mm},\cite{Caswell:1974gg},\cite{Egorian:1978zx}
respectively. The   
scheme-dependent  coefficients $\beta_2$ and $\beta_3$ 
are known from the symbolical calculations, which were done in  
\cite{Tarasov:1980au},\cite{Larin:1993tp} and  
\cite{vanRitbergen:1997va},\cite{Czakon:2004bu} correspondingly.

The solution of the RG-equation (\ref{beta}), namely 
\begin{eqnarray*}
\ln \frac{\mu^2}{M^2_q}=\int_
{a_s(\mu^2)}^{a_s(M^2_q)} \frac{dx}{\beta_0 x^2+\beta_1 x^3+\beta_2 x^4+\beta_3 x^5}
\end{eqnarray*}  
can be expressed through the $\ln$-dependent terms  $c_1$-$c_4$.  
They  have the following form 
\begin{eqnarray}
\nonumber 
c_1&=&\beta_0\ln\frac{\mu^2}{M^2_q}~, ~ c_2=\beta^2_0\ln^2\frac
{\mu^2}{M^2_q}+\beta_1\ln\frac{\mu^2}{M^2_q}~, \\ \nonumber 
c_3&=&\beta^3_0\ln^3\frac{\mu^2}{M^2_q}
+\frac{5}{2}\beta_0\beta_1\ln^2\frac{\mu^2}{M^2_q} 
+\beta_2\ln\frac{\mu^2}{M^2_q}~, \\ \nonumber 
c_4&=&\beta^4_0\ln^4\frac{\mu^2}{M^2_q}+\frac{13}{3}\beta_1\beta^2_0\ln^3
\frac{\mu^2}{M^2_q} \\ \nonumber 
&+&\frac{3}{2}(\beta^2_1+2\beta_0\beta_2)\ln^2\frac
{\mu^2}{M^2_q}+\beta_3\ln\frac{\mu^2}{M^2_q}        
\end{eqnarray}
We will also use the  four-loop approximation of the  anomalous mass  dimension in the 
$\rm{\overline{MS}}$-scheme, which is defined as 
\begin{eqnarray}
\label{gamma}
\gamma_m(a_s)&=&\mu^2\frac{\partial\ln(\overline{m}_q(\mu^2))}{\partial\mu^2}=-\sum\limits_{i=0}^{3} \gamma_i a_s^{i+1}~  
\end{eqnarray}
The numerical expressions of its coefficients in 
powers of $n_l$  have the following form  
\begin{eqnarray*}
&&\gamma_0=1~, \\
&&\gamma_1=4.0694-0.13888n_l~, \\
&&\gamma_2=17.204-2.3381n_l-0.02700n^2_l~, \\
&&\gamma_3=80.117-18.537n_l+0.2935n^2_l+0.00579n^3_l~.
\end{eqnarray*}
The first scheme-independent coefficient $\gamma_0$ was 
calculated in the works  \cite{Jones:1974mm},\cite{Caswell:1974gg}.
The presented above  $\rm{\overline{MS}}$-scheme results  for 
$\gamma_1$  and  $\gamma_2$ follow from the 
analytical calculations, performed in 
\cite{Tarrach:1980up},\cite{Nachtmann:1981zg} and 
\cite{Tarasov:1982gk}, \cite{Larinmass} respectively. The $n_l$-dependence  
for the term  $\gamma_4$ is obtained from its  analytical expression, 
simultaneously evaluated  in \cite{Chetyrkin:1997dh} and 
\cite{Vermaseren:1997fq}.
 
The solution of the RG-equation for  the running mass 
reads
\begin{eqnarray}
\label{mass}
\frac{\overline{m}_q(M^2_q)}{\overline{m}_q(\mu^2)}=
{\rm{exp}}\left(\int_{a_s(\mu^2)}^{a_s(M^2_q)}\frac{\gamma_m(x)dx}{\beta(x)}\right)
\end{eqnarray}
It can be expressed through the following terms 
\begin{eqnarray}
\nonumber 
b_1&=&\gamma_0\ln\frac{\mu^2}{M^2_q}~, \\ \nonumber
b_2&=&\frac{1}{2}\gamma_0(\gamma_0+\beta_0)
\ln^2\frac{\mu^2}{M^2_q}+\gamma_1\ln\frac{\mu^2}{M^2_q} \; , \\ 
\nonumber
b_3&=&\frac{1}{3}\gamma_0(\beta_0+\gamma_0)(\beta_0+\gamma_0/2)
\ln^3\frac{\mu^2}{M^2_q} \\ \nonumber 
&+&
\frac{1}{2}(\beta_1\gamma_0+2\gamma_1(\beta_0+\gamma_0))\ln^2\frac
{\mu^2}{M^2_q}+\gamma_2\ln\frac{\mu^2}{M^2_q}\; , \\ \nonumber
b_4&=&\frac{1}{24}(6\gamma_0\beta^3_0+11\gamma^2_0\beta^2_0+6\gamma^3_0
\beta_0+\gamma^4_0)\ln^4\frac{\mu^2}{M^2_q} \\ \nonumber
&+&\frac{1}{6}
(5\gamma_0\beta_0\beta_1+3\gamma^2_0\beta_1+3\gamma^2_0\gamma_1+
6\gamma_1\beta^2_0  \\ \nonumber 
&+& 9\gamma_0\gamma_1\beta_0)\ln^3\frac{\mu^2}{M^2_q}
+\frac{1}{2}(\gamma_0\beta_2+2\gamma_1\beta_1+3\gamma_2\beta_0  \\ \nonumber
&+& 2\gamma_0\gamma_2+\gamma^2_1)\ln^2\frac{\mu^2}{M^2_q}
+\gamma_3\ln\frac{\mu^2}{M^2_q}
\end{eqnarray}
Together with the derived above $c_1$-$c_4$ expressions, they  
contribute to the  $\ln$-dependent representation for the coefficients 
$l_1(\mu^2)$-$l_4(\mu^2)$ in the analogous to (\ref{l})  
relation  between pole and $\rm{\overline{MS}}$-scheme 
running heavy quark masses, normalized at the arbitrary 
renormalization scale:
\begin{eqnarray}
\nonumber 
l_1&=&b_1-z^{(1)}_m~, \\ \label{relations}
l_2&=&b_2-z^{(1)}_m(b_1+c_1)+(z^{(1)}_m)^2-z^{(2)}_m~, \\ \nonumber
l_3&=&b_3+((z^{(1)}_m)^2-z^{(2)}_m)(b_1+2c_1)\\ \nonumber 
&-&z^{(1)}_m(b_2+c_2+b_1c_1)-(z^{(1)}_m)^3+2z^{(1)}_m z^{(2)}_m-z^{(3)}_m~, \\ \nonumber
l_4&=&b_4-((z^{(1)}_m)^3-2z^{(1)}_m z^{(2)}_m+z^{(3)}_m)(b_1+3c_1) \\ \nonumber
&+&((z^{(1)}_m)^2-z^{(2)}_m)(b_2+2c_2+2b_1c_1+c^2_1) \\ \nonumber 
&-&z^{(1)}_m(b_3+c_3+b_2c_1+b_1c_2) \\ \nonumber 
&+&(z^{(1)}_m)^4- 3(z^{(1)}_m)^2z^{(2)}_m+2z^{(1)}_m z^{(3)}_m+(z^{(2)}_m)^2-z^{(4)}_m~.
\end{eqnarray}
The numerical expressions for the coefficients  
$z^{(i)}_m$ with $1\leq i\leq 4$ are presented in (\ref{num}),(\ref{num2}) 
and (\ref{num3}). 
The results for the terms  $z^{(40)}_m$ and $z^{(41)}_m$
with their  related  uncertainties  were obtained in the 
main part of the work with the help of mathematically rigorous OLS 
method and are given  in (\ref{answer+error}). 
Substituting now the presented above  expressions 
for the coefficients of the QCD RG-functions $\beta(a_s)$ and 
$\gamma_m(a_s)$ into the defined above  $\ln$-dependent terms
$c_1$-$c_4$ and $b_1$-$b_4$, 
which enter in the equations  (\ref{relations}),
we get  the following  
final expression for the 
the  flavour-dependent 
$\mathcal{O}(\alpha_s^4)$ explicit  relation between pole and $\rm{\overline{MS}}$-scheme  running  masses of heavy quarks at the arbitrary renormalization 
scale $\mu^2$:
\begin{eqnarray}
\nonumber
&&M_q=\overline{m}_q(\mu^2)\bigg[1+\bigg(\frac{4}{3}+\ln\frac{\mu^2}{M^2_q}\bigg) a_s(\mu^2)  \\ \nonumber 
&& +\bigg(16.110-1.0413n_l+(8.8471-0.36109n_l)\ln\frac{\mu^2}{M^2_q}  \\ 
\nonumber
&& +(1.79166-0.083333n_l)\ln^2\frac{\mu^2}{M^2_q} \bigg)a^2_s(\mu^2) \\
&&+ \bigg(239.29-29.700n_l+0.6526n^2_l \\ \nonumber 
&& +(129.408-15.3671n_l+0.31996n^2_l)\ln\frac{\mu^2}{M^2_q}  \\ \nonumber 
&& + (32.1047-3.05309n_l+0.060179n^2_l)\ln^2\frac{\mu^2}{M^2_q}  \\ \nonumber 
&&+(3.68286-0.370369n_l+0.0092591n^2_l)\ln^3\frac{\mu^2}{M^2_q}\bigg)
a^3_s(\mu^2) \\ \label{d4}
&&+\bigg(4457.6\pm 62.0-(864.25\pm 15.20)n_l+46.309n^2_l \\ \nonumber  
&&-0.6781n^3_l+(2463.69-449.525n_l+22.6677n^2_l \\ \nonumber
&&-0.31920n^3_l)\ln\frac{\mu^2}{M^2_q}+(651.105-113.1880n_l  \\ \nonumber
&&+5.58420n^2_l-0.079958n^3_l)\ln^2\frac{\mu^2}{M^2_q}+(102.7368 \\ \nonumber
&&-15.70710n_l+0.732188n^2_l-0.0100295n^3_l)\ln^3\frac{\mu^2}{M^2_q} \\ \nonumber 
&&+(8.05625-1.270540n_l+0.0665503n^2_l \\ \nonumber 
&&-0.00115739n^3_l)\ln^4\frac{\mu^2}{M^2_q}\bigg)a^4_s(\mu^2)+\mathcal{O}(a^5_s) \bigg]
\end{eqnarray}
At fixed $n_l=3,4,5$ the expressions of the 
terms, contributing to  $l_4$ 
agree with the results of \cite{Marquard:2015qpa}.\\

{\bf Note added}\\

After two previous versions of this work were submitted for publication 
the detailed description of the  numerical calculations  
of \cite{Marquard:2015qpa} appeared in \cite{Marquard:2016dcn}, where 
the authors clarified the number of questions, raised in this  
work and in the talk \cite{Kataev:2016jai}.  
The new work \cite{Marquard:2016dcn} presented 21 more 
precise numerical expressions  for $z_m^{(4)}$-coefficient 
at fixed $n_l$ values, which vary in the region  $0\leq n_l\leq
20$.  For these data set the OLS  method gives the 
more precise analogs of the 
terms, obtained above   in (\ref{answer+error})  from three given in   
\cite{Marquard:2016dcn} terms only, namely  
$z^{(40)}_m=-3654.14\pm 0.76$, $z^{(41)}_m=756.94\pm 0.07$. 
Their  central values coincide with the presented in  \cite{Marquard:2016dcn}
results of the  diagram-by- diagram calculations. 
One can see that as the result of taking extra 18 numerical 
expressions, given in \cite{Marquard:2016dcn}
the  uncertainties of the  
OLS method are  decreasing significantly and are becoming more physical. 
The more precise OLS  values 
of the terms, determined above in (\ref{l_4K}) and (\ref{d4}) using three 
equations only, read  $l_4=(3567.6\pm 0.76)-(745.72\pm 0.07)n_l+43.396n^2_l-0.6781n^3_l$ and $d_4$=$4469.04\pm 0.76$-$(864.14\pm 0.07)n_l$+46.307$n^2_l$-0.6781$n^3_l$. The presented OLS-based expression for $l_4$ is in the excellent agreement 
with the result of numerical diagram-by-diagram calculations, given in 
\cite{Marquard:2016dcn}. The central values of the two OLS-determined terms 
also agree rather well with similar values of the same terms  
from (\ref{l_4K}) and (\ref{d4}), obtained by us in  the main part of this work.
The agreement of the results obtained in this work by means of    
mathematically consistent  OLS method   
with the results of diagram-by-diagram calculations , which were  presented 
only recently in \cite{Marquard:2016dcn}, should be considered as the argument 
in favour of self-consistency  of the performed at supercomputer Lomonosov 
of MSU complicated and important  numerical calculations, described in 
\cite{Marquard:2015qpa}, \cite{Marquard:2016dcn}, and may be used in 
various physical and mathematical studies.

\end{document}